\documentclass{article}
\usepackage{spconf,amsmath,graphicx}

\usepackage[usenames,dvipsnames]{color}
\usepackage{verbatim}
\usepackage{balance}
\usepackage{subfig}
\usepackage{algpseudocode}
\usepackage{multirow}
\newsavebox{\measurebox}
\usepackage{minibox}
\usepackage{array,multirow,booktabs}
\usepackage{graphicx}
\usepackage{siunitx}
\usepackage{amsfonts,amssymb}
\newcommand{\squeezeup}{\vspace{-2.5mm}}

\title{Improving Maximum Likelihood Difference Scaling method to measure inter content scale}

\name{\parbox{\linewidth}{\centering Andréas Pastor$^{\star}$ \qquad Lukáš Krasula$^{\dagger}$ \qquad
Xiaoqing Zhu$^{\dagger}$ \qquad Zhi Li$^{\dagger}$ \qquad Patrick Le Callet$^{\star}$}}
\address{$^{\star}$ Nantes Université, Ecole Centrale Nantes, CNRS, LS2N, UMR 6004, F-44000 Nantes, France \\
        $^{\dagger}$Netflix Inc., Los Gatos, CA, USA}
\begin{document}
%
\maketitle
\begin{abstract}

The goal of most subjective studies is to place a set of stimuli on a perceptual scale. This is mostly done directly by rating, e.g. using single or double stimulus methodologies, or indirectly by ranking or pairwise comparison. All these methods estimate the perceptual magnitudes of the stimuli on a scale. However, procedures such as Maximum Likelihood Difference Scaling (MLDS) have shown that considering perceptual distances can bring benefits in terms of discriminatory power, observers' cognitive load, and the number of trials required. One of the disadvantages of the MLDS method is that the perceptual scales obtained for stimuli created from different source content are generally not comparable. In this paper, we propose an extension of the MLDS method that ensures inter-content comparability of the results and shows its usefulness especially in the presence of observer errors.
\end{abstract}

\squeezeup
\section{Introduction}
\label{sec:intro}
\squeezeup


In general, it is important to perform subjective tests to validate systems (e.g. coding algorithms), benchmark objectives metrics, or create datasets to train machine learning or deep learning models. 
However, collecting these annotations can be time-consuming and expensive to perform. Due to the subjectivity of the data, we don't always have an agreement in the judgment of people and accurate estimation is important to reduce noise and uncertainty on collected data. In that case, to boost these subjective tests is essential to allocate annotation resources on the right stimuli. 

Multiple methodologies exist to rate stimuli, with direct estimation like Absolute Category Rating (ACR) or Double Stimuli Impairment Scale (DSIS), or indirectly, two-Alternative Forced Choice (2AFC) or pairwise comparison (PC). PC is a more reliable method because observers only need to provide their preference on each pair, comparisons are more sensitive, which is important to improve the \textit{discriminability}.

PC experiments are generating matrices of values indicating how many times a stimulus has been preferred over another. These Pair Comparison Matrix (PCM) can be translated to a continuous scale, using models (e.g. Thurstone\cite{Thurstone}, Bradley and Terry \cite{BradleyTerry}, Tversky \cite{tversky1972elimination}). Due to the pairwise manner of presenting stimuli, the PCM size, and the number of possible comparisons is growing quadratically with the number of stimuli, introducing \textit{efficiency} in a subjective protocol. A lot of previous works have focused on \textit{active sampling} solutions \cite{glickman2005adaptive, pfeiffer2012adaptive, li2012analysis, li2013boosting, chen2013pairwise, ye2014active} and more recently \cite{hybrid_mst, xu2018hodgerank, simpson2020scalable, mikhailiuk2021active, demers2021active, men2021subjective, boostingPCforQA} to select the most informative pairs, and minimize experimental effort while maintaining accurate estimations.

In this work, we want to present an extension to the MLDS methodology \cite{maloney2003maximum, knoblauch2008mlds}. MLDS can be used to separately estimate scales of different stimuli sets with intra-set supra-threshold perceptual differences, see in section 2.1. But it isn't possible to estimate a global scale where the stimuli of different sets can be related, section2.2. These sets of stimuli from different contents are compared using different sampling strategies, introduced in section 3. With simulations, objectives are to demonstrate that our propositions can achieve good \textit{discriminability} over content with \textit{robustness} against bad annotator behavior and with \textit{efficiency}.

\begin{figure}[!htbp]
\centering
 \mbox{ \parbox{1\textwidth}{
  \begin{minipage}[b]{0.5\textwidth}
  {\label{fig:img_trial_example}\includegraphics[width=\textwidth]{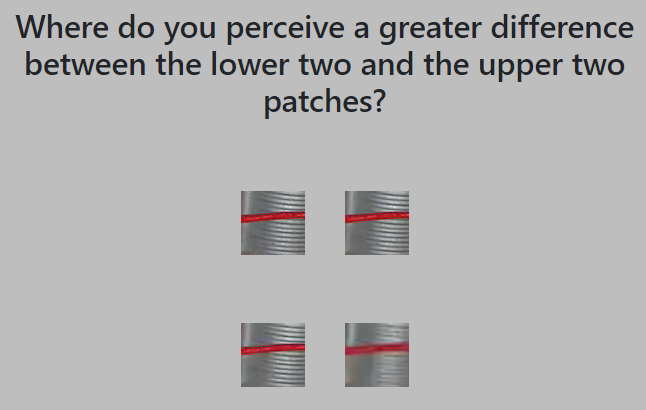}}
  \end{minipage}
   }}
\caption{Example of intra-content trial presented to an observer to rate where he perceives a greater difference in the quadruplet.}
\squeezeup  
\squeezeup  
\label{fig:trial_example}
\end{figure}

\squeezeup  
\squeezeup
\section{Maximum Likelihood Difference Scaling}
\label{sec:mlds}
\squeezeup

In this section, we explain the principle of MLDS\cite{maloney2003maximum, knoblauch2008mlds} and its extension to estimate the scale of cross-content perception.

The content $C_{i}$ consists of a reference stimulus $S^{i}_1$ and $n - 1$ modified versions of this reference stimuli: $S^{i}_2, ..., S^{i}_n$. This set of stimuli $C_{i} = \{ S^{i}_{1}, S^{i}_{2}, ..., S^{i}_{n} \}$ is pre-ordered along a physical continuum, with the assumption that larger alteration introduce higher perceptual difference:
\begin{equation}
    S^{i}_{1} <  S^{i}_{2} <  ... <  S^{i}_{n}
\end{equation}

The observer is presented during a trial with a quadruplet of stimuli $(S_{i}, S_{j}, S_{k}, S_{l})$, see figure \ref{fig:trial_example} for an example, and asked to estimate where the perceptual distance is greater: inside the pair $(S_{i}, S_{j})$ or $(S_{k}, S_{l})$. A trial outcome is either 0 or 1, corresponding to the judgment of 
\begin{equation}
\left | S_{i} - S_{j} \right | - \left | S_{k} - S_{l} \right | > 0.
\end{equation}

During solving, MLDS estimates scalar values $(\phi^{i}_{1}, ..., \phi^{i}_{n})$ to predict the observer judgment in each quadruplet.

\squeezeup
\subsection{Intra-content difference scaling}

This system of equations can be reduced to a set of linear equations with the previous assumption and solved with a General Linear Model (GLM), using a link function $\eta = logit(\pi(x))$ or $probit(\pi(x))$ where $\pi(x)$ is $\mathbb{P}(Y = 1|X_{1} = x_{1}, . . . , X_{n} = x_{n})$ and,
\begin{equation}
\pi(x) = F(\phi_{1}X_{1} + \phi_{2}X_{2} + . . . + \phi_{n}X_{n})
\end{equation}

where F is the inverse of the link function $\eta$. An example for 5 quadruplets of a content with 7 stimuli (1-7):
\begin{equation}
    \begin{pmatrix}
    2 & 4 & 5 & 6 \\ 
    1 & 2 & 3 & 7 \\  
    1 & 5 & 6 & 7 \\  
    1 & 2 & 4 & 6 \\  
    3 & 5 & 6 & 7 
    \end{pmatrix}
\end{equation}

yield the following matrix,
\begin{equation}
    X = \begin{pmatrix}
    0 & 1 & 0 & -1 & -1 & 1 & 0 \\ 
    1 & -1 & -1 & 0 & 0 & 0 & 1 \\
    1 & 0 & 0 & 0 & -1 & -1 & 1 \\
    1 & -1 & 0 & -1 & 0 & 1 & 0 \\
    0 & 0 & 1 & 0 & -1 & -1 & 1
    \end{pmatrix}
\end{equation}

Once the $(\phi^{i}_{1}, ..., \phi^{i}_{n})$ are estimated, they can be plotted as a perceptual curve, for example one of the line in fig. \ref{fig:glm_solve}. 

\squeezeup
\subsection{Inter-contents difference scaling}

In the previous section, we introduced the MLDS method and how it estimates perceptual differences inside a set of stimuli from a content. However, this method can't directly relate and scale the difference of perception of a group of contents.

To estimate a scaling factor for each perceptual curve, one perceptual curve per content, we added inter-content comparisons. Where a quadruplet is composed of a pair of stimuli from a content $C_{i}$ and a pair from a content $C_{j}$: $(S^{i}_{a}, S^{i}_{b}, S^{j}_{c}, S^{j}_{d})$. The observer is asked, similarly as in intra-content comparison, to judge where he perceives a large perceptual distance:

\squeezeup
\begin{equation}
\left | S^{i}_{a} - S^{i}_{b} \right | - \left | S^{j}_{c} - S^{j}_{d} \right | > 0
\end{equation}

\begin{figure}[!htbp]
\centering
 \mbox{ \parbox{1\textwidth}{
  \begin{minipage}[b]{0.5\textwidth}
  {\label{fig:img_glm_solve}\includegraphics[width=\textwidth]{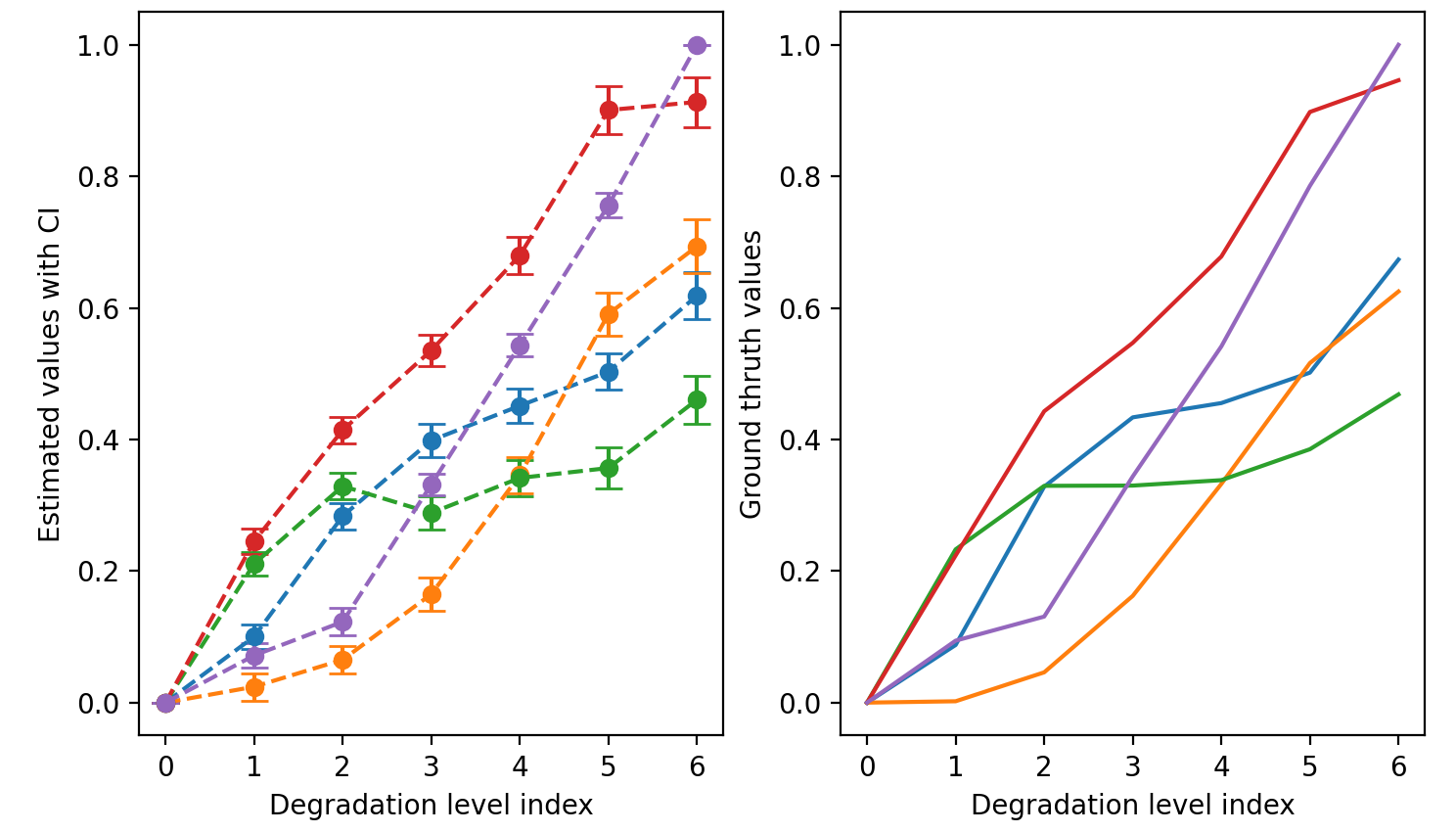}}
  \end{minipage}
   }}
\caption{Example of 5 contents with 7 stimuli each and their perceptual curves, estimated using GLM and different scaling. On the right, the ground truth data to recover, and on the left estimations produced with simulated annotations of 15 participants, CIs are obtained via bootstrapping \cite{bootstrap_ci}.}
\squeezeup  
\squeezeup  
\label{fig:glm_solve}
\end{figure}

With the addition of this set of inter-content trials, we can construct a larger matrix $X$, solved with GLM, and estimate the difference of scales between a group of perceptual curves, figure \ref{fig:glm_solve}.
\squeezeup
\begin{equation}
    X = \begin{pmatrix}
          & \begin{matrix} 0 & 0 & 0 & 0 & 0 & 0 \end{matrix} \\
    X_{1} & \begin{matrix} 0 & 0 & 0 & 0 & 0 & 0 \end{matrix} \\
          & \begin{matrix} 0 & 0 & 0 & 0 & 0 & 0 \end{matrix} \\
    \begin{matrix} 0 & 0 & 0 & 0 & 0 & 0 \end{matrix} &       \\ 
    \begin{matrix} 0 & 0 & 0 & 0 & 0 & 0 \end{matrix} & X_{2} \\ 
    \begin{matrix} 0 & 0 & 0 & 0 & 0 & 0 \end{matrix} &       \\ 
    \begin{matrix} 1 & -1& 0 & 0 & 0 & 0 \end{matrix} & 
    \begin{matrix} -1 & 1& 0 & 0 & 0 & 0 \end{matrix} \\
    \begin{matrix} 1 & 0 & -1 & 0 & 0 & 0 \end{matrix} & 
    \begin{matrix} 0 & 0 & -1 & 0 & 1 & 0 \end{matrix} \\
    \end{pmatrix}
\label{X12_eq}
\end{equation}

$X$ is composed of submatrices $(X_{1},X_{2})$ of intra-content comparisons for a content $C_{1}$ and $C_{2}$ and a third matrix of inter-content comparisons (e.g. last 2 rows in the example above).

\squeezeup 

\squeezeup
\section{Experimental Simulations}
\label{sec:exp_setup}

In this section, we will experiment with the modification that we added to the MLDS methodology and check how we can select relevant quadruplets to minimize the experimental effort and be robust to spammer behavior. All comparisons are performed with simulated annotations.

\squeezeup
\subsection{Quadruplet selection}
\subsubsection{Intra-content quadruplet selection}

The quadruplets for intra-content comparisons are generated following the recommendation of MLDS paper \cite{maloney2003maximum, knoblauch2008mlds}, where there are $\binom{L}{4}$ quadruplets for a L levels difference scaling experiment. In our case, we selected $L=7$ levels to be estimated (a reference stimuli + 6 distorted versions) yielding 35 quadruplets to evaluate for each content.

\subsubsection{Inter-content quadruplet selection}

In the case of inter-content comparisons, we can create for a first content $M = \frac{L \times (L - 1)}{2}$ pairs to be compared with the $M$ pairs of a second content, which is not tractable when the number of contents is large: if N is the number of content, there are $\frac{N \times (N - 1)}{2} \times M^{2}$ possible quadruplets. This will be the baseline \textbf{\textit{fulldesign}} that we will improve from in the next section.

Next, we describe some sampling strategies to reduce the fraction of comparisons to perform and how we validate them in simulations.

\squeezeup
\subsection{Sampling strategies}
\subsubsection{Similar alteration levels comparisons}

To reduce the amount of inter-content quadruplets to evaluate, we decided to compare only the same alteration levels pair in inter-content where a quadruplet is formed as follows $(S^{i}_{a}, S^{i}_{b}, S^{j}_{a}, S^{j}_{b})$, with $a$ and $b$ 2 different alteration levels.
On top of that to reduce the number of pairs available in comparison we constrained in a case named \textbf{\textit{"consecutive design"}} to have $b = a + 1$ and in a second case \textbf{\textit{"reference design"}} $a = 1$ where $S^{i}_{1}$ and $S^{j}_{1}$ are the references.

with these 2 cases, and the associated design constraints, we reduce the number of comparisons to $L - 1$ quadruplets: where the $L - 1$ pairs, created from $L$ levels of alterations, of a content A are compared once to the $L - 1$ pairs of a content B. Then when comparing N contents together we get $(L - 1) \times \frac{N \times (N - 1)}{2}$ quadruplets.

\squeezeup
\subsubsection{k-connection contents comparisons}

The second part we can improve on is to reduce comparisons to only in sets of k contents, instead of the N-1 other available contents:
$(L - 1) \times (N \times k)$ comparisons with k the number of compared contents.

The k-connection is performed over all contents $A = {A_{1}, ..., A_{n}}$ by connecting a content $A_{i}$ to the set of contents $(A_{i+1}, A_{i+2}, ..., A_{i+k})$ with a modulo operation $(i + k) \% n$, to loop back on the first contents if needed for the last $k-1$ contents. We experimented with different values of k and report the most interesting one in the result section.

\squeezeup
\subsubsection{AFAD: Adaptive Far Apart Discard design}

When performing selecting comparisons, it is better to focus on close enough stimuli to avoid comparing perceptual distances that have a large difference, see eq.\ref{eq_large_dist}: if we present these pairs to an observer, at best it will result in the same judgment as to all other previous observers, or in the worse case (e.g. spammer or error behavior) be an inverse judgment that will mislead the solving algorithm.

\begin{equation}
    (S_{i} - S_{j}) >> (S_{k} - S_{l})
\label{eq_large_dist}
\end{equation}

To avoid these cases, we will discard far apart quadruplets using the estimation of $\phi^{i}_{j}$ values. After each set of annotations from an observer, our difference scaling implementation fits and estimates the difference in the stimuli, then discards the bottom 20\% of quadruplets with the largest difference scale. This strategy is reported as \textbf{\textit{"AFAD"}} in the results section.

\squeezeup
\squeezeup
\section{Simulations results}
\squeezeup

\begin{figure}[!htbp]
\centering
 \mbox{ \parbox{1\textwidth}{
  \begin{minipage}[b]{0.5\textwidth}
  {\label{fig:img_result25_05_log}\includegraphics[width=\textwidth]{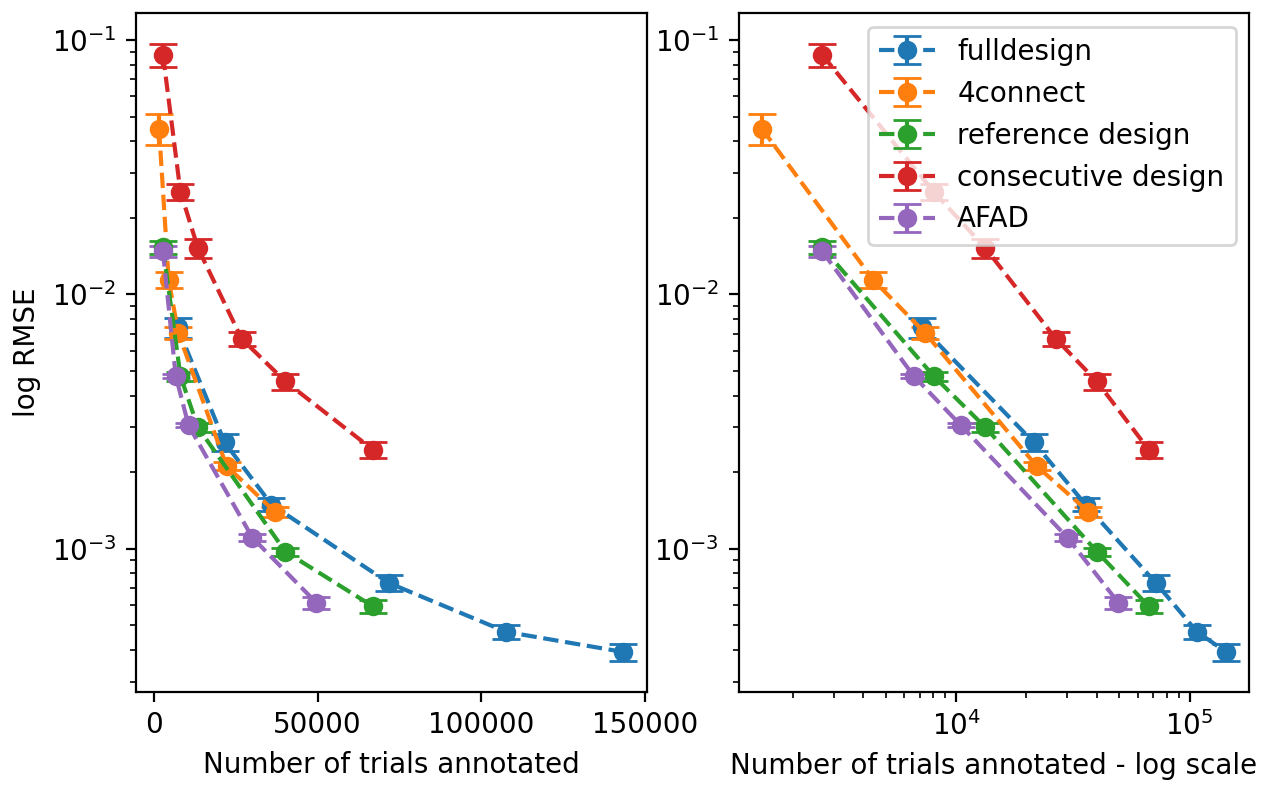}}
  \end{minipage}
   }}
\caption{RMSE results for 25 contents with 7 alteration levels: X axis log scale RMSE (left: number of annotations, right: number of annotations - log scale) with a 5\% probability of inverting an annotation vote.}
\squeezeup  
\label{fig:result_05_RMSE}
\end{figure}
\squeezeup
\squeezeup
\begin{figure}[!htbp]
\centering
 \mbox{ \parbox{1\textwidth}{
  \begin{minipage}[b]{0.5\textwidth}
  {\label{fig:img_result_cc}\includegraphics[width=\textwidth]{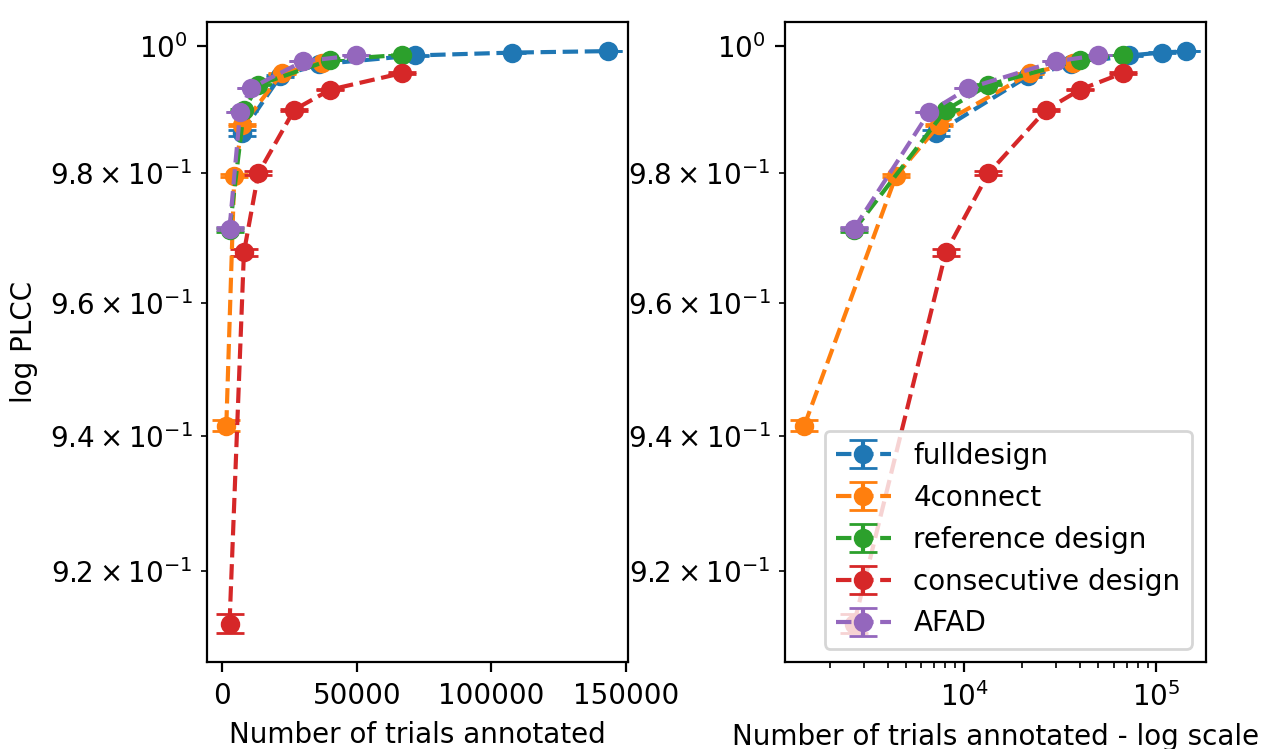}}
  \end{minipage}
   }}
\caption{Pearson correlation results for 25 contents with 7 alteration levels: X axis log scale Pearson correlation (left: number of annotators, right: number of annotations) with a 5\% probability of inverting an annotation vote.}
\squeezeup  
\label{fig:result_05_cc}
\end{figure}

To select and compare the performances of the different quadruple selection methodologies, we compared them through simulation.
To evaluate the improvement of the proposed method, a Monte Carlo simulation was conducted. 25 contents with 7 levels of alterations were designed whose scores were randomly selected using the following principle. To construct the scores of each alteration in a content, we uniformly sample on $[0, 5/6]$ 6 values and add a first value of 0 corresponding to the reference. 
Then we compute the cumulative sum over this array of 7 elements to yield the final 7 subjective scores to retrieve through simulation for a content, starting at 0 and ranging until a maximum of 5. An example of designed contents can be checked in figure \ref{fig:glm_solve}, right part.

The following assumptions were made for simulation: 1) each stimulus has a single score, described above; 2) in each observation, the observed judgment of a quadruplet follows a Gaussian random variable, X, $N(|ma - mb| - |mc - md|, 1)$, with mi, 4 scalars corresponding to the scores of the stimuli in the quadruplet and a fix standard deviation of 1 to model stochasticity of the participant perception; 3) each observer has a 5\% probability to produce a mistake, i.e., inverting his judgment, to test the robustness of the solution against outlier behavior; and 4) each trial is independent. Raw trial judgments are obtained by taking the sign of the sampled value from X and mapping it to the values 0 or 1.

The simulations use our python implementation of the original difference scaling algorithm, with GLM to convert raw trials judgment to estimated scores. 
RMSE and Pearson correlation are reported between the estimated scores and the designed ones. Each of the simulations was run at least 100 times to report 95\% confidence intervals.

\squeezeup 

\begin{figure}[!htbp]
\centering
 \mbox{ \parbox{1\textwidth}{
  \begin{minipage}[b]{0.5\textwidth}
  {\label{fig:img_result25_20_log}\includegraphics[width=\textwidth]{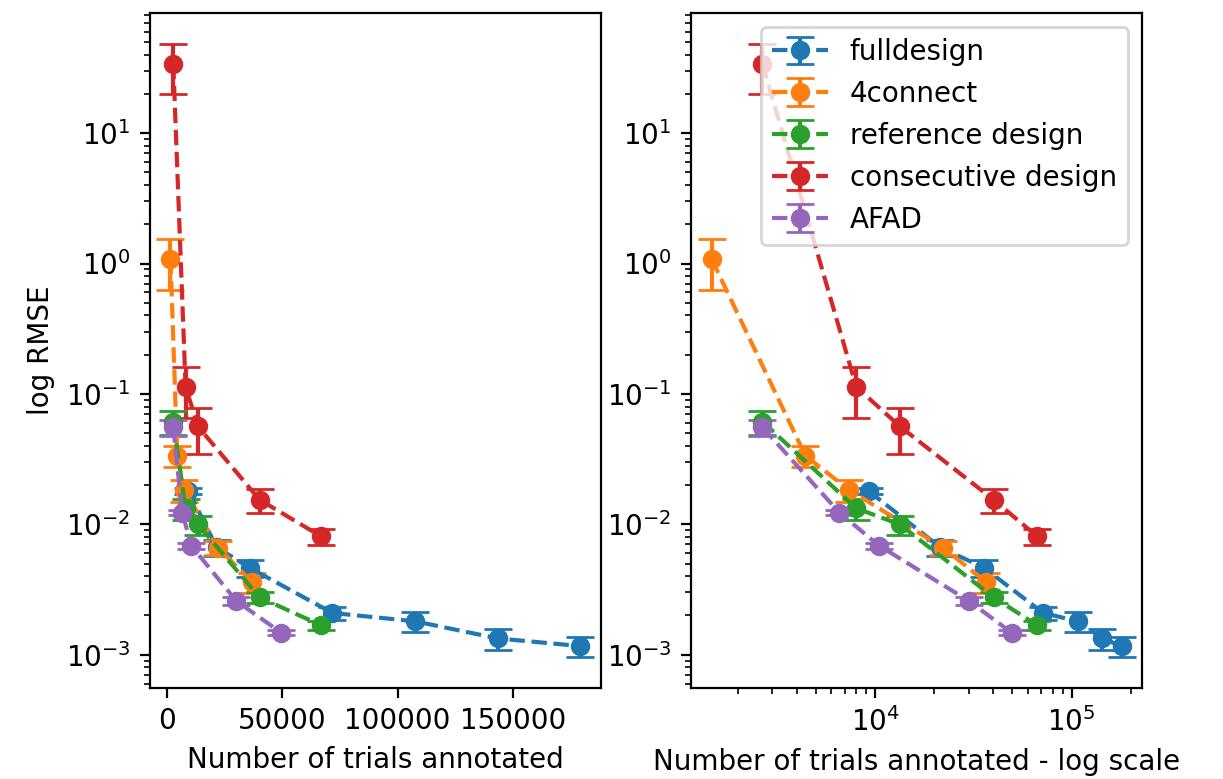}}
  \end{minipage}
   }}
\caption{RMSE results for 25 contents with 7 alteration levels: X axis log scale RMSE (left: number of annotations, right: number of annotations - log scale) with a 20\% probability of inverting an annotation vote.}
\squeezeup  
\label{fig:result_20_RMSE}
\end{figure}

\begin{figure}[!htbp]
\centering
 \mbox{ \parbox{1\textwidth}{
  \begin{minipage}[b]{0.5\textwidth}
  {\label{fig:img_result_cc20}\includegraphics[width=\textwidth]{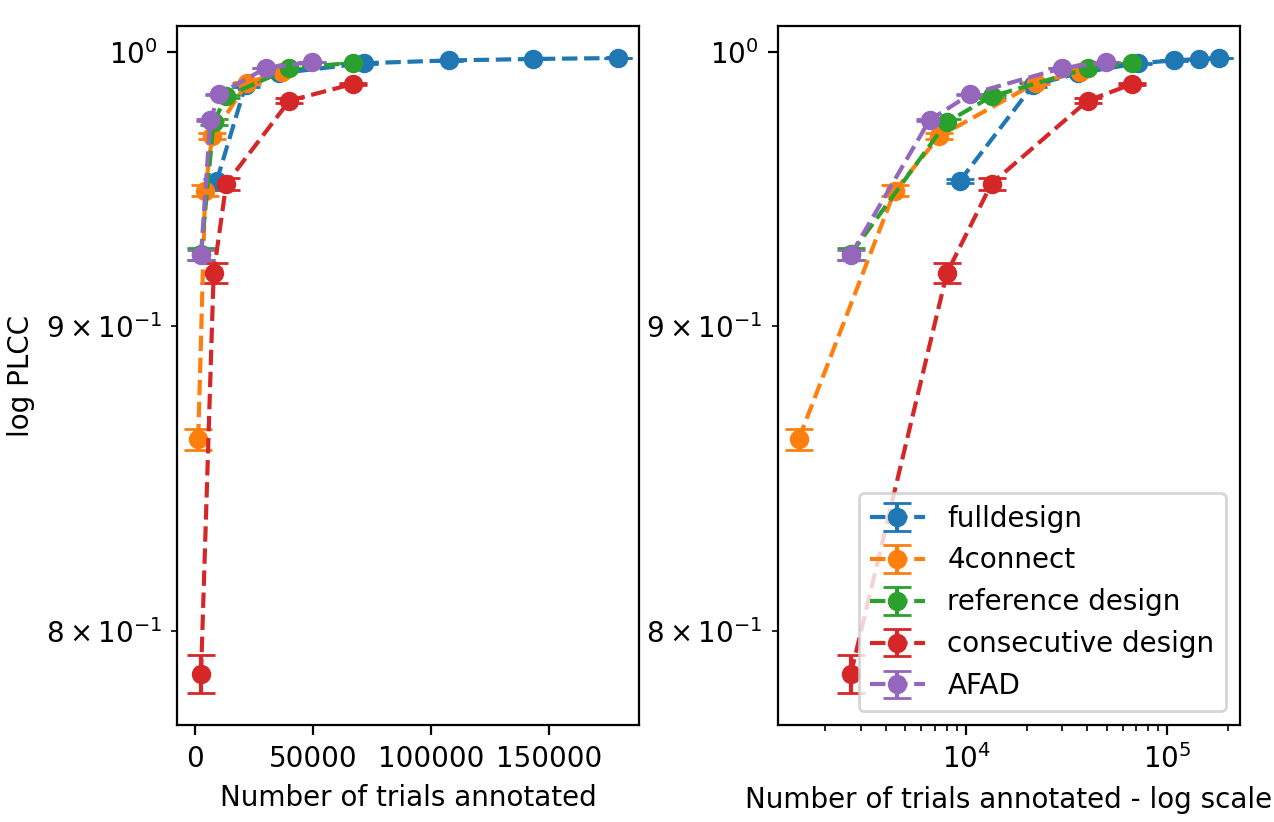}}
  \end{minipage}
   }}
\caption{Pearson correlation results for 25 contents with 7 alteration levels: X axis log scale Pearson correlation (left: number of annotators, right: number of annotations - log scale) with a 20\% probability of inverting an annotation vote.}
\squeezeup 
\label{fig:result_20_cc}
\end{figure}

In figures \ref{fig:result_05_RMSE} and \ref{fig:result_05_cc}, we present the simulation results for the different sampling strategies that we investigated. It shows that for a fixed number of trials (e.g $10^{4}$), the original \textbf{\textit{fulldesign}} performs worse than some designs and especially \textbf{\textit{AFAD}}. The proposed design converges faster to smaller estimation errors and a higher correlation score.
To estimate the number of annotations saved by one design over another, we linearly fitted the data:
\begin{equation}
    log(numberAnnotations) = A \times log(RMSE) + B
\end{equation}
We use coefficients A and B to predict the number of annotations needed for RMSE scores of $R = [0.01, ..., 0.1]$. Finally, we compute the percentage of annotators saved for each RMSE score in $R$ and average them. In table \ref{tab:res_dist_05}, we read that AFAD design needs 39.7\% less annotations compared to full design to achieve the same RMSE value on average.

\begin{table}[!htbp]
\centering
\begin{tabular}{|c|c|} 
 \hline
        & full design \\
\hline
4-connect design & -0.68\% \\
reference design & -26.67\% \\
consecutive design & 228\% \\
\textbf{AFAD} & -39.7\%  \\
 \hline
\end{tabular}
\caption{ Distance between curves in percent.}
\label{tab:res_dist_05}
\end{table}

\squeezeup
\squeezeup
\squeezeup
\subsection{Observer Error Resiliency}

When we increase the ratio of annotation error from 5\% to 20\% in fig. \ref{fig:result_20_RMSE}-\ref{fig:result_20_cc}, the proposed solution \textbf{\textit{AFAD}} still estimates better than the other methods. We can see as well that the estimation error is higher than the one report in fig. \ref{fig:result_05_RMSE}-\ref{fig:result_05_cc}, especially with a small number of annotations.

\squeezeup
\section{Conclusion and Future Work}
\label{sec:conclusion}
\squeezeup

As we emphasized previously, MLDS methodology has a limitation when we need to compare perceptual scales of different source stimuli. The extension that we introduce and the \textit{active sampling} strategies that we suggested, is providing a solution to perform a subjective experiment in this paradigm of supra-threshold estimation. The strategy presents robustness to bad annotator behavior and can effectively reduce the number of comparisons to perform.

This work could be applied to collect robustly and effectively, in a large-scale crowdsourcing experiment, human judgments on distorted video patches, via video coding algorithms.



\newpage
\balance
\small{
\bibliographystyle{IEEEbib}
\bibliography{main}
}
\end{document}